\documentclass[twocolumn,secnumarabic, graphics,floatfix,prl]{revtex4-1}

\usepackage{amsfonts}
\usepackage[T1]{fontenc}
\usepackage[ansinew]{inputenc}
\usepackage{graphicx} %%For loading graphic files
\usepackage{amsmath}
\usepackage{latexsym}
\usepackage{amssymb}
\usepackage{color}
\usepackage[normalem]{ulem}
\usepackage{epstopdf}
\usepackage{float}
\usepackage{textcomp} %degresC

\newcommand{\Ser}{\color{black}}

\newcommand{\tnd}{\rm{tn}\delta}%{\rm{tan}\delta}%
\newcommand{\dgam}{\dot \lambda}

\definecolor{linkcolor}{rgb}{0,0,0.6} %hyperlink
\usepackage[pdftex,colorlinks=true, pdfstartview=FitV, linkcolor= linkcolor, citecolor= linkcolor, urlcolor= linkcolor, hyperindex=true,hyperfigures=true]{hyperref} %hyperlink% or

\begin{document}

%\title{Memory effects in the stress and in the dielectric susceptibility   \\
	%of a polymer glass stretched at constant strain rate}
\title{Simultaneous memory effects in the stress and in the dielectric susceptibility \\
	of a stretched polymer glass}

\author{J. Hem$^1$,   C.  Crauste-Thibierge$^1$, F. Cl\'ement$^2$, D.R.  Long$^2$, S. Ciliberto$^1$,}
\affiliation{$1$ Univ of Lyon, Ens de Lyon, Univ Claude Bernard, CNRS, Laboratoire de Physique, UMR 5672, F-69342 Lyon, France}
%\affiliation{$1$ Laboratoire de Physique, CNRS UMR5672,  Universit\'e de Lyon, \'Ecole Normale Sup\'erieure, 
%	46 All\'ee d'Italie, 69364 Lyon, France}
\affiliation{$2$ Laboratoire Polym\`eres  et Mat\'eriaux Avanc\'es, CNRS/Solvay, UMR 5268, 69192 Saint Fons Cedex, France}

\begin{abstract}
We report experimental evidence that a polymer stretched at constant strain rate $\dgam$ presents complex memory effects after that $\dgam$ is set to zero at a specific strain $\lambda_w$ for a duration $t_w$, ranging from $100$s to $ 2.2\times10^5$s. When the strain rate is resumed, both the stress and the dielectric constant relax to the unperturbed state non monotonically. The relaxations depend on the observable, on $\lambda_w$ and on $t_w$. Relaxation master curves are obtained by scaling the  time and the amplitudes by $\ln (t_w)$. The dielectric evolution also captures the distribution of the relaxation times, so the results impose strong constraints on the relaxation models of polymers under stress and they can be useful for a better understanding of memory effects in other disorder materials. 
\end{abstract}
\maketitle
%%%%%%
%-----------------------------------------------------------------------------------------------------------------------------
%\section{ Introduction }
%-----------------------------------------------------------------------------------------------------------------------------

Aging of amorphous and heterogeneous materials is characterized by a slow relaxation toward a new state after that a perturbation has been applied \cite{Struick_Book_1978}. This relaxation usually spans several orders of magnitude in time and may present memory effects, which depend on the sample history (see Ref. \cite{Nagel_RMP_2019} for a recent review). 

%Memory effects are ubiquitous of aging and they have been observed in the physical properties of many different disordered systems such as the magnetic response of spin glasses \cite{Bouchaud_2005,Castillo_PRL_2002}, the mechanical \cite{Kovacs_1963} and dielectric properties of polymer glasses \cite{Kovacs_dielectric_1997,Bellon_2000,Bellon_2002}, frictional interfaces \cite{Fineberg_memory_PRL_2006,frictional_interface_PRL_2018}, the load on crumpled thin sheets \cite{Nagel_2002_Sheet,memory_sheet_PRL_2017}, the electrical conductance of several materials \cite{Amir_PRL_2011}, the stress in gels \cite{gel_memory_2014} and the properties of granular systems \cite{Granular_Memory_1995,Trizac_granular_aging_PRL_2014,vibrated_granular_PRL_2000,granular_PRL_2020}.

Memory effects have been observed in glassy and disordered materials \cite{Bouchaud_2005,Vincent_2007,Castillo_PRL_2002,Kovacs_1963,Vincent_1960,Kramer_1970,Ricco1985,Ricco1990,Yee1988,G'Sell1992,Kovacs_dielectric_1997,Bellon_2000,Bellon_2002,Amir_PRL_2011,gel_memory_2014,Zapas1974}, and also in many other disordered systems \cite{Fineberg_memory_PRL_2006,frictional_interface_PRL_2018,Nagel_2002_Sheet,memory_sheet_PRL_2017,Granular_Memory_1995,Trizac_granular_aging_PRL_2014,vibrated_granular_PRL_2000,granular_PRL_2020}.{\Ser Memory effects may take the form of a Kovacs-like nonmonotonic relaxation, in which the response of a state variable, to an external perturbation, evolves in one direction before turning around at a timescale and an amplitude which depend on the sample history. These Kovacs-like relaxations have been observed in the mechanical \cite{Kovacs_1963,Vincent_1960,Kramer_1970,Ricco1985,Ricco1990,Yee1988,G'Sell1992} and electrical \cite{Kovacs_dielectric_1997} responses of polymer glasses and melts \cite{Zapas1974}, in the static coefficient of friction \cite{Fineberg_memory_PRL_2006, frictional_interface_PRL_2018}, in the load on crumpled thin sheets \cite{Nagel_2002_Sheet,memory_sheet_PRL_2017} and in the mechanical response of granular systems \cite{Granular_Memory_1995,vibrated_granular_PRL_2000,Trizac_granular_aging_PRL_2014,granular_PRL_2020}.}

Several theoretical models \cite{McKenna_2003,Bertin_JPA_2003,Sibani_PRE_2010,Amir_PNAS_2012,Long_theory_Kovacs_2006,Long_theory_Kovacs_2018} have been developed to explain, with common underlying physical principles, the similar features of these complex relaxations observed in very different systems. One of the key assumptions of these models is the existence into these disordered systems of a distribution of relaxation times (DRT) which is used to explain the origin of aging and of non-monotonic relaxations. However it is hard to discern between the different descriptions using experiments in which only one observable is measured. Thus to give more insight into this problem, we performed experiments on a stretched polymer glass where one gets quantitative information on the memory of several observables measured simultaneously and on the evolution of the DRT.

In this letter we describe these experiments and report the observation that a stretched polymer glass exhibits complex memory effects of both the stress and of the dielectric susceptibility. Specifically, we show that, after a perturbation of the strain rate,  the relaxation dynamics of the stress and the dielectric susceptibility present non monotonic responses which depend not only on the sample history, as in the above mentioned Kovacs-like memory effect, but also on the observable. %\cite{Kovacs_1963,Vincent_1960,Kramer_1970,Kovacs_dielectric_1997,Zapas1974,Fineberg_memory_PRL_2006, frictional_interface_PRL_2018,Nagel_2002_Sheet,memory_sheet_PRL_2017,Granular_Memory_1995,vibrated_granular_PRL_2000,Trizac_granular_aging_PRL_2014,granular_PRL_2020}, but also on the observable. 
%{\Jer In this letter, we investigate stress aging experiments and observe that stretched polymer glasses exhibit complex memory effects of both the stress and the dielectric susceptibility. In the experiment, the polymer is successively stretched at constant strain rate, held at fix deformation, and finally stretched again with the same initial strain rate. It was shown as early in the 1960s \cite{Vincent_1960,Kramer_1970} that the stress response when recovering is non monotonic and depend on the sample history, such as the waiting time, as in the above mentioned Kovacs-like memory effect. Here we show that after perturbation of the strain rate, the relaxation dynamic of the dielectric susceptibility also present overshoot and memory effects but significantly delayed compare to the stress response, so that two observables may present different relaxations.} 
Furthermore from the evolution of the dielectric spectra, we get insights on the evolution of the polymer DRT. {\Ser Thus our results impose constraints for modeling not only the universal features of memory effects but also the polymer relaxation dynamics under deformation, which is widely studied, both experimentally \cite{Loo_SCI_2000,Lee_JCP_2008,Lee_SCI_2009,Ediger2009,Lee_MAC_2010,Perez_MAC_2016,Sahli_PRM_2020,Kalfus_MAC_2012} and theoretically \cite{robbins2009,schweizer2004,chen2009b,conca2017,riggleman2007,McKenna_2003}. }

In our experiment the investigated polymer is an extruded film MAKROFOL DE\textsuperscript{\textregistered} 1-1 000000 from BAYER based on Makrolon\textsuperscript{\textregistered} polycarbonate (PC) of thickness 125~$\mu$m. The films are cut into rectangular pieces of size $42\times21cm$ with a dog bone shape before being tied to the tensile machine. During the experiment, we measure the instantaneous stress $\sigma(t)$ and deformation $\lambda(t)=L(t)/L_0$ of the films. The experiments are performed at imposed $\dgam\simeq 10^{-4}$s and at $T=25^\circ$C much below the glass transition temperature $T_g= 150^\circ$C.{\Ser In parallel to the stress measurement we use the dielectric spectroscopy to investigate the dynamics of the relaxation processes at work in the stretched PC film.\cite{PC_Type,Kremer2003}} Dielectric spectroscopy is indeed a convenient way to probe in-situ and without destruction the relaxation times of stretched polymers \cite{Kalfus_MAC_2012,Perez_MAC_2016,Sahli_PRM_2020}. It is worth to mention that other techniques, such as NMR measurements \cite{Loo_SCI_2000} or probe molecule experiments \cite{Lee_JCP_2008,Lee_SCI_2009}, can be used but dielectric spectroscopy is very versatile. In the rest of the paper, we will focus only on the loss tangent $\tnd(f) =-\epsilon''(f)/\epsilon'(f)$, which does not depend on the film thickness, being the ratio of the imaginary ($\epsilon''$) and real ($\epsilon'$) parts of the dielectric susceptibility, at frequency $f$.{\Ser For the dielectric measure of our samples, the film is sandwiched between two electrodes and conductive gel layers which insure a uniform electrical contact. The electrodes are connected to a homemade dielectric spectrometer, which measures the dielectric spectrum every $32$s in the frequency range $[10^{-1}-10^{3}]$Hz, simultaneously on all of the frequencies and with an accuracy of a few percent \cite{Perez_RSI_2015}. Thus it allows us to follow the evolution of the dielectric spectra during the stretching protocols which lasts more than 2 hours. In order to get the required repeatability we average the results over at least 3 tests performed in different samples submitted to the same stretching protocol. More details about the experimental apparatus can be found in Ref. \cite{Perez_RSI_2015,Perez_MAC_2016,Sahli_PRM_2020}.}

%spanning simultaneously $4$ decades in frequency in the range $[10^{-1}-10^{3}]$Hz in a time interval of 32s.   }   
%{\Jer Data are acquired at $8192pts/s$ in segments of $64s$ which are further processed by Welch algorithm to get the spectra. Note that an overlap of $50\%$ between each segments is considered to better follow the change in the sample during the experiment, so that the effective increment is $32s$. The results are averaged over $3-4$ tests to ensure the repeatability}

\begin{figure}[h]
	\centering  
	\includegraphics[width=1\linewidth]{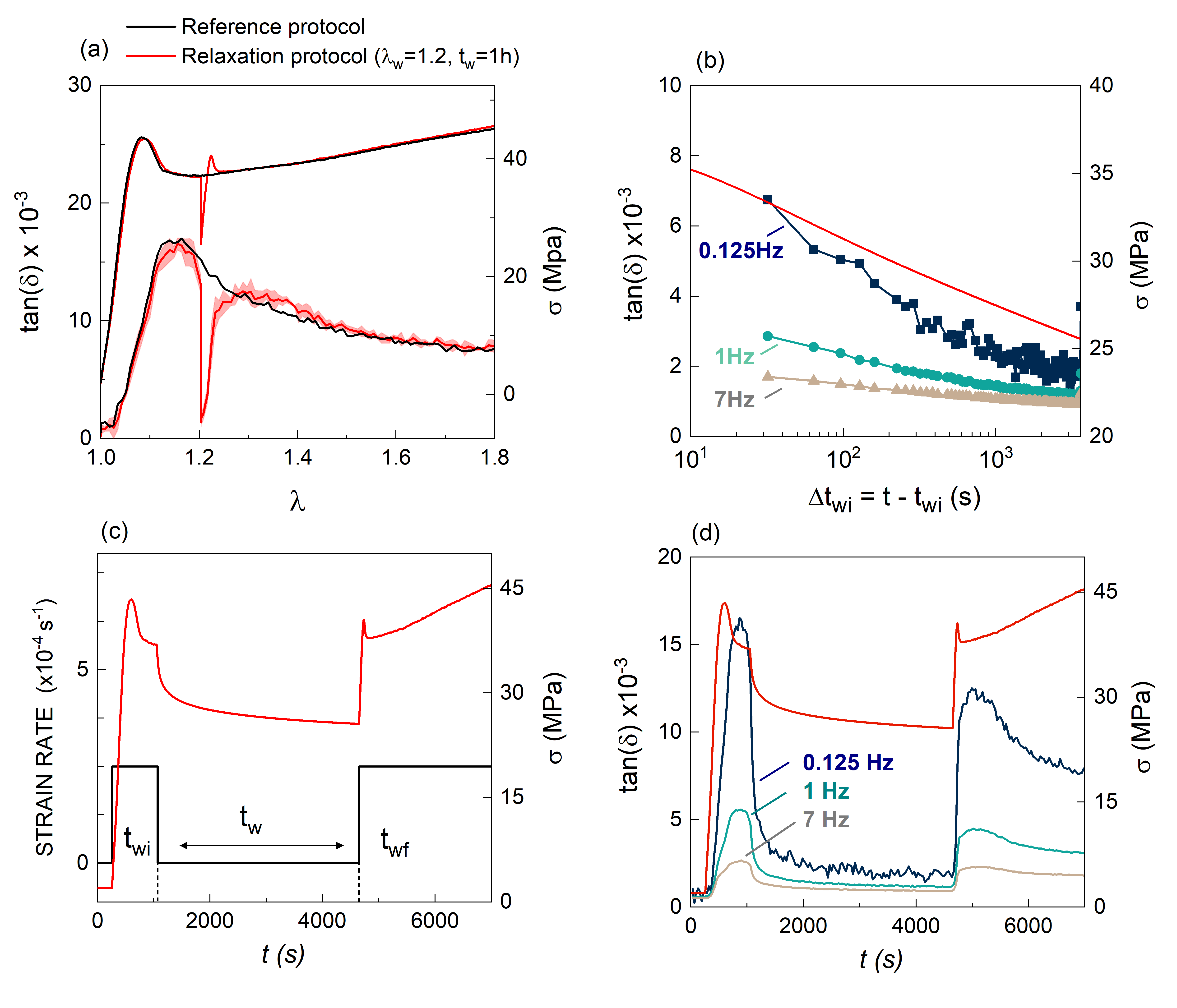}
	\caption{ (a) Dielectric susceptibility $\tnd$ at $f=0.125Hz$ and nominal stress $\sigma=F/S_o$ with $S_o=210\times0.125mm^2$ as function of the strain $\lambda$. Black lines: constant strain rate protocol $\dgam=2.5\times10^{-4}s^{-1}$, called reference curves. Red lines: relaxation protocol $\lambda_w = 1.2$ and $t_w = 1h$. {\Ser The red line thickness corresponds to  one standard deviation.}(b) Temporal evolution $\tnd$ at $f=0.125, 1, 7Hz$ and $\sigma$ during the waiting step over $t_w=1h$.(c) $\dgam(t)$ and $\sigma(t)$ during the relaxation protocol with $\lambda_w = 1.2$, $t_w = t_{wf} - t_{wi} = 1h$. Times $t_{wi}$ and $t_{wf}$ delimit the beginning and the end of the waiting interval respectively.(d) $\tnd(t)$ at $f=0.125, 1, 7Hz$ and $\sigma(t)$.   	
	}
	\label{Fig:Protocol}
\end{figure}

We first recall the behavior of the stress and the dielectric susceptibility as a function of $\lambda$ measured at constant strain rate $\dgam$, that we have discussed in Ref. \cite{Perez_MAC_2016,Sahli_PRM_2020}. Fig.\ref{Fig:Protocol}(a) depicts the stress evolution for polycarbonate at $\dgam=2.5\times10^{-4}s^{-1}$ (black curve). The stress first increases until the yield stress at about $\lambda = 1.08$, then, successively decreases, a regime known as strain softening, reaches a plateau at about $\lambda = 1.20$ and finally increases again, which is known as strain hardening. The breaking occurs at about $\lambda = 1.8-2$ \cite{necking_note}. {\Ser The dielectric spectra in the range $[10^{-1}- 10^{3}]$Hz have been measured every $32$s, during which $\lambda$ has a negligible increase of less than $0.8\%$ at this $\dgam$.} As described in Ref. \cite{Kalfus_MAC_2012,Perez_MAC_2016,Sahli_PRM_2020}, dielectric measurements give access to the evolution of the dominant relaxation time $\tau_{\alpha}$ of the polymer by identifying the lower part of the dielectric spectrum as the high frequency tail of the $\alpha$ relaxation peak. In particular, we have shown in Ref. \cite{Sahli_PRM_2020} that the increase (decrease) of the dielectric susceptibility in the range $[10^{-1}- 10^{3}]$Hz reflects a decrease (increase) of $\tau_\alpha$, which depends on $\lambda$ and the strain rate $\dgam$. Therefore, the evolution of $\tnd$ at $f=0.125Hz$ depicted in Fig.\ref{Fig:Protocol}(a) shows an acceleration of the relaxation times until the softening/onset of plastic flow (maximum around $\lambda=1.15-1.17$) and a slow down during the strain hardening regime that we have studied in Ref. \cite{Sahli_PRM_2020}. The results of the measurements performed at constant $\dgam$ (black curves in Fig.\ref{Fig:Protocol}(a)) will be called ''reference curves''.    

To investigate the memory effect, we have modified the protocol at constant $\dgam$ in order to include an aging step at a specific $\lambda$ in the post yield regime of the film. More precisely, the film is first stretched at $\dgam=2.5\times10^{-4}s^{-1}$ then held at the waiting strain $\lambda_{w}$ during the waiting time $t_w$, and finally further stretched with the same initial $\dgam$. As an example for the case $\lambda_w=1.2$ (plastic flow) and $t_w=1h$, the evolutions of $\sigma$ and of $\tnd(0.125)$ are illustrated as a function of $\lambda$ in Fig.\ref{Fig:Protocol}(a). The corresponding time evolution is illustrated in Figs.\ref{Fig:Protocol}(c,d) where the instantaneous strain rate $\dgam$ is indicated too. Three different intervals can be distinguished: During the initial one ($\lambda<\lambda_w$ and $t<t_{wi}$), $\sigma$ and $\tnd$ are identical to the reference curves, see Fig.\ref{Fig:Protocol}(a). During the waiting interval (i.e.$\lambda_w=1.2$ and $t_{wi} \le t \le t_{wf}$), they decrease logarithmically with time, which is the signature of an aging process, see Fig.\ref{Fig:Protocol}(b). Finally as soon as $\dgam$ is resumed for $t \ge t_{wf}$, they increase again. In Fig.\ref{Fig:Protocol}(a), the comparison with the reference curves obtained with the constant $\dgam$ protocol shows that $\sigma$ and $\tnd(0.125)$ finally recover their reference curves. However notice that, in spite of the fact that $\sigma$ and $\tnd(0.125)$ have a similar dependence on $\lambda$ before the aging step, the recoveries of the two observables are rather different, see Fig.\ref{Fig:Protocol}(a). In addition, we shall see that these recovery processes also depend on $t_w$ and on $\lambda_w$.

%-----------------------------------------------------------------------------------------------------------------------------
%\section{ Discussion on the relaxation times & Comments on Fig 2 }
%-----------------------------------------------------------------------------------------------------------------------------

\begin{figure}[h]
	\centering  
	\includegraphics[width=1\linewidth]{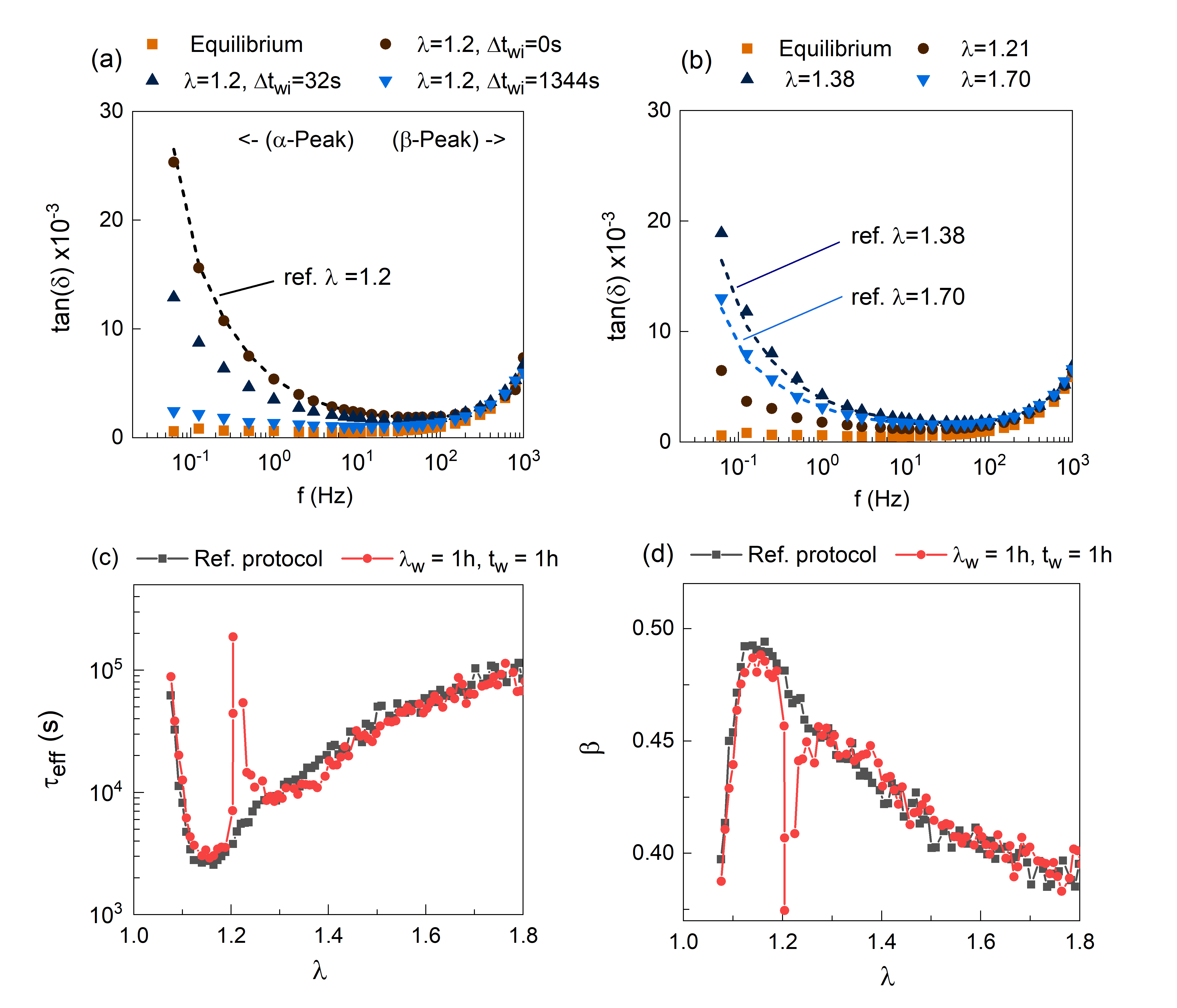}
	\caption{Dielectric spectra $\tnd$ recorded at different times for the relaxation protocol at $\lambda_w=1.2$ and $t_w=1h$.(a) Evolution during the aging interval at $\lambda=1.2$ with $\Delta t_{wi}=t-t_{wi}$ and (b) during the recovery interval for $\lambda>1.2$. Dashed lines correspond to the reference spectra recorded at constant $\dgam$. The evolutions of $\tau_{eff}$ and $\beta$ are plotted in (c) and (d) respectively. Black line: constant $\dgam$ protocol. Red line: relaxation protocol.
	}
	\label{Fig:TauAndBeta}
\end{figure}

Before presenting these results, we point out that Fig.\ref{Fig:Protocol}(d) shows that the response of $\tnd$ to $\dgam$ increases by reducing the frequency $f$. From this frequency dependence we get useful information on the DRT evolution by studying the whole spectrum of $\tnd$. Thus we plot the $\tnd$ spectra in Fig.\ref{Fig:TauAndBeta}(a) during the waiting interval at $\lambda_w = 1.2$ and in Fig.\ref{Fig:TauAndBeta}(b) during the recovery for $t>t_{wf}$. We first notice that the spectra evolve in a significant way only for $f<100$Hz. Furthermore, we see in Fig.\ref{Fig:TauAndBeta}(a) that the spectrum at $\lambda=1.2$, at the beginning of the aging ($t=t_{wi}$), is identical to the reference one obtained during the constant $\dgam$ protocol. Then, during the waiting interval for increasing $\Delta t_{wi}=t-t_{wi}$, the spectrum progressively decreases towards the equilibrium spectrum (i.e. the spectrum at $\lambda=1$). Inversely, in Fig.\ref{Fig:TauAndBeta}(b), the spectra during the recovery interval increase again until they reach the reference spectrum around $\lambda = 1.3$, i.e. $\Delta t=t-t_{wf}=400$s.

%From this frequency dependence we get  useful information on the  DRT evolution by studying the whole spectrum of $\tnd$. %, that most theoretical models try to predict. 
The main relaxation time can be estimated from these dielectric spectra following the method presented in Ref.\cite{Sahli_PRM_2020} in which we have shown that at low frequencies $\tnd (f) \simeq (\sin{\beta\pi/2}) /(2 \pi\  f \ \tau_{eff})^\beta$ where the characteristic time $\tau_{eff}$ and the exponent $\beta$ have to be determined. Within the Cole-Cole \cite{havriliak1966} model, $\tau_{eff}$ is  proportional to the main relaxation time $\tau_\alpha$ of the polymer \cite{tau_alpha} and the exponent $\beta$ is related to the width of the distribution of the relaxation times, specifically the smallest is $\beta<1$ the larger is the distribution \cite{Lee_MAC_2010,Sahli_PRM_2020}. Thus the fit of the low frequencies part of the spectra in Figs.\ref{Fig:TauAndBeta}(a),\ref{Fig:TauAndBeta}(b) (recorded at different times) allows us to estimate the evolution of the DRT during the relaxation protocol. In particular, we see in Figs.\ref{Fig:TauAndBeta}(c),\ref{Fig:TauAndBeta}(d) that, during the waiting interval, $\tau_{eff}$ ($\beta$) increases (decreases), meaning that not only the dynamics slow down but that the distribution of times is broadening. When the strain rate is resumed the dynamics accelerate again and the distribution shrinks, recovering the reference state after the resumption. Hence, this is the evidence that the DRT depends on the driving history and that $\dgam$ accelerates the polymer dynamics, even if the acceleration is not monotonous in $\lambda$.{\Ser The results of fig.\ref{Fig:TauAndBeta} strengthen quantitatively early observations on  DRT reported in Ref.\cite{Ediger2009}.} 

%-----------------------------------------------------------------------------------------------------------------------------
%\section{ Comments on Fig 3 }
%-----------------------------------------------------------------------------------------------------------------------------

\begin{figure}[h]
	\centering  
	\includegraphics[width=1\linewidth]{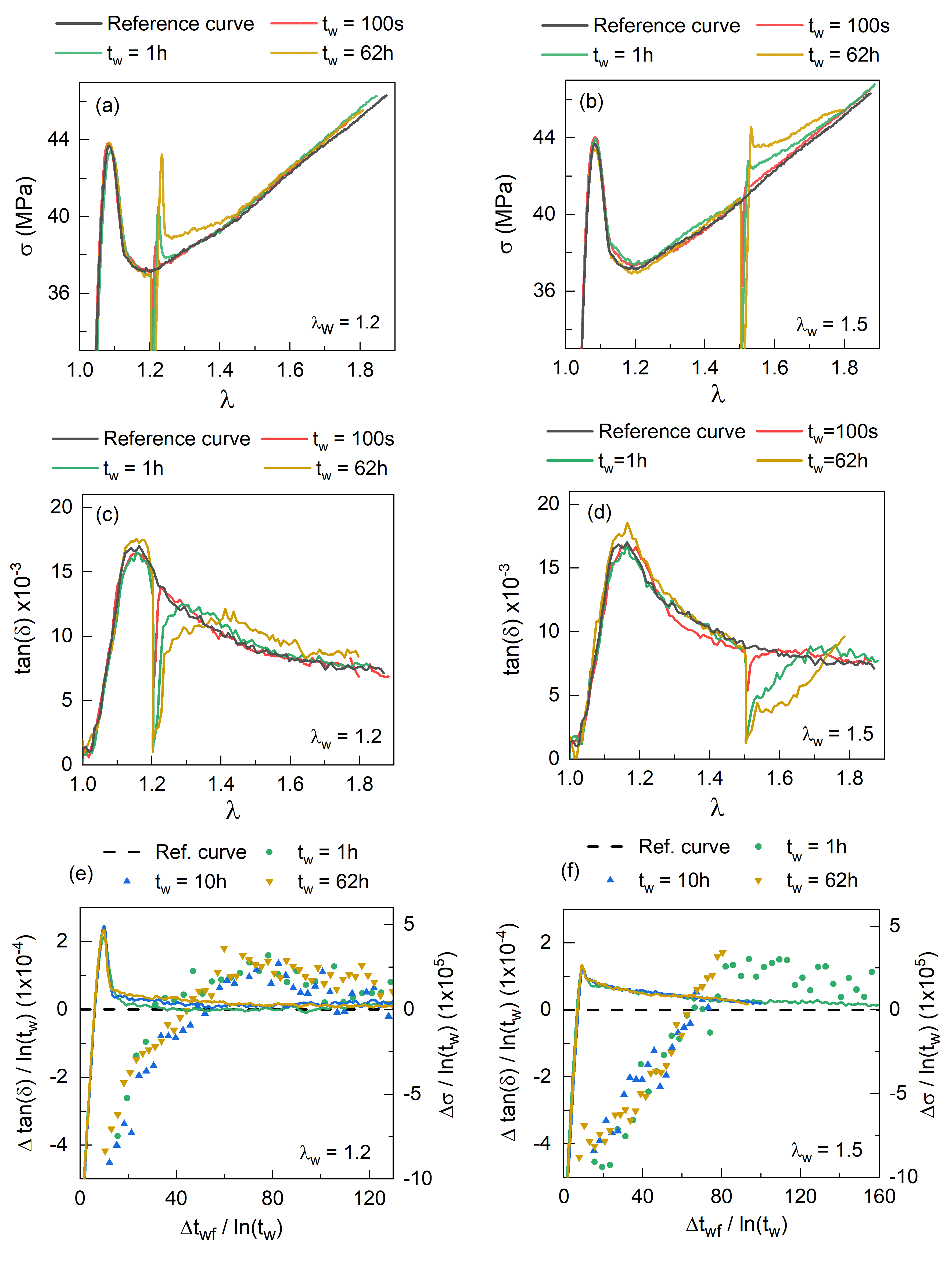}
	\caption{
	{\Ser (a-d) Stress $\sigma$ and dielectric susceptibility $\tnd$ at $f=0.125$Hz as a function of $\lambda$ for different $\lambda_w$ and $t_w$. Curves obtained by averaging 3 tests with $t_w=62$h and 4 tests with $t_w=100$s and $1$h. (a,c) $\sigma$ and $\tnd$ at $\lambda_w=1.2$ (in the plastic flow regime). (b,d) $\sigma$ and $\tnd$ at $\lambda_w=1.5$ (in the strain hardening). The reference curves are indicated in black. Note the stress axis have been expanded in (a,b) with respect to Fig.\ref{Fig:Protocol}(a). (e,f) Evolution of $\Delta \sigma / \ln(t_w)$ (continuous lines) and $\Delta \tnd/ \ln(t_w)$ (symbols) as function of $\Delta t / \ln(t_{w})$ for  $t \ge t_{wf}$, (e) at $\lambda_w = 1.2$ and (f) at $\lambda_w = 1.5$ (see text for details).
%		with $\Delta \sigma=\sigma-\sigma_{ref}$, $\Delta \tnd=\tnd-\tnd_{ref}$ and $\Delta t = t-t_{wf}$   	
	}
	}
	\label{Fig:MainResults}
\end{figure}

To study the evolution of $\tnd$ as a function of $t_w$ and $\lambda_w$, we now select $\tnd$ measured at $f=0.125$Hz as a typical amplitude of the low frequency spectra, which depend on $\tau_\alpha$ and $\beta$. Figs.\ref{Fig:MainResults}(a-f) bring more details on the recovery processes by showing different results of tests performed with different waiting times and waiting strains $\lambda_w = 1.2$ (plastic flow) in Figs.\ref{Fig:MainResults}(a,c) and $\lambda_w = 1.5$ (strain hardening) in Figs.\ref{Fig:MainResults}(b,d).
 When $\dgam$ is resumed at $t_{wf}$, we observe that the relaxations of  $\sigma$ and  $\tnd$ are both non-monotonic and exhibit an overshoot above the reference curve before relaxing to it, with different shapes of the evolutions of the two observables. 
% When $\dgam$ is resumed at $t_{wf}$, we observe that the relaxation of the stress $\sigma$ is non-monotonic, as in previous measurements \cite{Vincent_1960,Kramer_1970},an  overshoot above the reference curve before relaxing to it. The same  with different shapes of the evolutions of the two observables. 
 On one side, the stress overshoot has a distinct and large amplitude while occurring rapidly after the resumption, see Figs.\ref{Fig:MainResults}(a,b). In contrast, the overshoot of $\tnd$ has a weaker amplitude and occurs much later in time, see Figs.\ref{Fig:MainResults}(c,d). 
%Note that these differences can not be a problem of time resolution in the Fourier calculation process since identical phenomenology has been also observed with shorter temporal window.

Moreover, we find that the waiting time has a strong impact on the relaxation. In the case of $\sigma$, Figs.\ref{Fig:MainResults}(a,b) show that, as long as $t_w$ increases, the overshoot peak shifts progressively to longer time and presents a higher and sharper amplitude. In the case of $\tnd$, Figs.\ref{Fig:MainResults}(c,d) show that an increase of $t_w$ also shifts the peak time and slows down the relaxation to the reference curve. Also in the dielectric case, we observe that a longer $t_w$ affect the peak amplitude. 

Besides the waiting time, the choice of the waiting strain also modifies the relaxation. Figs.\ref{Fig:MainResults}(a,b)show qualitatively that the stress overshoot peak is higher and narrower at $\lambda_w=1.2$ than at $\lambda_w=1.5$ and at the same time Figs.\ref{Fig:MainResults}(c,d) show that $\tnd$ recovers much faster the reference curve at $\lambda_w=1.2$ than at $\lambda_w=1.5$. The difference is significant. For example at $t_{w} = 1h$ and $\lambda_{w}=1.2$, $\tnd$ recovers the reference curve at about $\lambda=1.45$, equivalent to $\Delta t=t-t_{wf}=1024s$ after the the resumption of the stretching. This is much faster than for the case at $t_{w} = 1h$ and $\lambda_{w}=1.5$ for which the recovery is even too long to occur during our experiment ($\lambda > 1.85$ equivalent to $\Delta t > 1240s$). {\Ser In particular, this enhancement of the mobility recovery delay with $\lambda_w$ fully supports the long one observed in multicreep experiments \cite{Ediger2009}, in which the aging relaxation is performed in the strain hardening regime.}
% At this point of the discussion, we want to mention that an increase of $\lambda_w$ also correlates with an increase of the stress level in the aging step so that, it is not clear yet whether the fundamental parameter is the deformation regime ($\lambda_w$), the aging stress \cite{Kramer_1970} or both.}

%\sout{Since the relaxation of the observables are affected by the waiting time $t_w$ at fixed waiting strain $\lambda_w$ and reciprocally by $\lambda_{w}$ at fixed $t_{w}$, one must conclude that these two aging parameters $\lambda_w$ and $t_w$ govern independently the relaxation during the recovery process.}

%-----------------------------------------------------------------------------------------------------------------------------
%\section{ Comments on Fig 4 }
%-----------------------------------------------------------------------------------------------------------------------------

%\begin{figure}[!h]
	%\centering  
	%\includegraphics[width=1\linewidth]{Graph_Fig4.png}
	%\caption{ 
	%}      	
	%\label{Fig:Overshoot}
%\end{figure}

To be more precise regarding the dependence of the relaxations on those aging parameters we analyze the difference between the stress curve and the reference one $\Delta \sigma=\sigma-\sigma_{ref}$ as a function of the  time $\Delta t=t-t_{wf}$ elapsed after resumption. Figs.\ref{Fig:MainResults}(e,f) show that all the relaxation curves, measured after different waiting times $t_w$, can be superimposed {\Ser on a master curve} by scaling  $\Delta \sigma$ and $\Delta t$ by $\ln (t_{w})$, for $t_w>100$s. Moreover, we observe that this relation holds regardless of the waiting strain value ($\lambda_w = 1.2$ or $\lambda_w = 1.5$). As a consequence, this means that the peak time and the peak amplitude depend on $\ln(t_w)$. Regarding the dependence of the peak properties on $t_w$ and $\lambda_w$, we note, that the peak amplitude $\Delta \sigma_p$ is different for the two waiting strains, being higher at $\lambda_w = 1.2$ than at $\lambda_w = 1.5$ for all $t_w$. In contrast the peak time $\Delta t_{p}$ is independent on the waiting strain, see Fig.\ref{Fig:MainResults}(e,f).  
	
Finally, it is interesting to see that this way of scaling the data also works for the dielectric susceptibility. In a similar way than before, we have studied the evolution of the difference between the $\tnd$ curve at $f = 0.125Hz$ and the reference one $\Delta \tnd=\tnd-\tnd_{ref}$ as function of $\Delta t$. Figs.\ref{Fig:MainResults}(e,f) show that all the relaxation curves merge {\Ser on a master curve} by rescaling $\Delta \tnd$ and $\Delta t$ by $\ln(t_{w})$. This is observed again for the two waiting strains investigated, either at $\lambda_w = 1.2$ (Fig.\ref{Fig:MainResults}(c)) or at $\lambda_w = 1.5$ (Fig.\ref{Fig:MainResults}(d)). Note that for the dielectric case, the peak time at $\lambda_w = 1.5$ is delayed compared to $\lambda_w = 1.2$ which was not the case for the stress relaxation. The logarithmic dependence on $t_w$ shows, in particular, that the memory of the aging history can also occur in $\tnd$, so that we provide evidence that the memory appears simultaneously in different observables but in different ways.

Concerning the scaling $\Delta t/\ln(t_w)$, we point out that in several systems with non-monotonous relaxations, the memory effects present a $\Delta t/t_w$ scaling \cite{Nagel_RMP_2019,Amir_PNAS_2012,memory_sheet_PRL_2017}, which however is not generic. In several cases \cite{Vincent_2007}, a heuristic form of scaling $\Delta t/t_w^\mu$ is used, which  has been theoretically justified \cite{Sibani_PRE_2010}. If $\mu<<1$ then $\ln t_w$ cannot be distinguished from $t_w^\mu$, and our scaling $\Delta t/\ln(t_w)$ can be understood within this theoretical framework. In these models the DRT does not depend on time whereas our measurements show that it does evolve, thus our measure of the DRT evolution will help in constructing more precise models. It remains to understand why the time scale of the relaxations are different for $\sigma$ and $\tnd$.    
%It is interesting to compare our results with those experiments in which the dependence  of the mechanical response of polymers  as a function of  the time  $\tilde t_w$ spent  in the glassy state after a temperature quench from the melt. In such a case it is observed that  the  amplitude of the main yield stress peak also undergoes a similar logarithmic dependence on  $\tilde t_w$ \cite{Roth_Book_2017}. However in contrast to our experiment the positions of the yield peak does not  depend $\tilde t_w$. 
%This difference could be related to the way the waiting step is realized.
% When the polymer ages after a temperature quench, experiments are usually performed at constant pressure whereas in our experiment, the polymer ages from the plastic regime at constant strain.  From this perspective, it must be concluded that stretching  affects substantially the aging and the recovery processes. 
It is interesting to notice that for polymers the amplitude of the main yield stress peak also undergoes a similar logarithmic dependence on the aging time $\tilde t_w$, which in this case is the time spent in the glassy state after a temperature quench from the melt \cite{Bauwenscrowet_1982,Meijer_2005,Klompen_2005,Roth_Book_2017}. However in contrast to our experiment the positions of the yield peak does not depend on $\tilde t_w$.  
% This difference could be related to the way the waiting step is realized. When the polymer ages after a temperature quench, experiments are usually performed at constant pressure whereas in our experiment, the polymer ages from the plastic regime at constant strain. 
This difference could be related to the way the waiting step is realized. Indeed when the polymer ages after a temperature quench no stress is applied to the sample before the measure of the loading curve, whereas in our experiment the polymer ages from the plastic regime at constant strain.
From this perspective, it must be concluded that stretching affects substantially the aging and the recovery processes, which is confirmed by the dependence of the memory effects on $\lambda$ observed in our experiment. The evolution of the dielectric susceptibility and of the DRT when a stress is applied after the time $\tilde t_w$ has never been measured, and no comparison with our results can be done. This comparison would be very useful for giving new insights to the polymer relaxations under stress.    

{\Ser As a conclusion, we have shown that a stretched polymer presents complex memory effects in the non-monotonous relaxations  of $\sigma$ and $\tnd$.  Although non monotonous response of $\sigma$ have been already observed in stretched polymers \cite{Vincent_1960,Kramer_1970,Ricco1985,Ricco1990,Yee1988,G'Sell1992} it is interesting to notice that $\tnd$ behaves in a similar way as a function of $\lambda_w$ and $t_w$. Non monotonous relaxation of the dielectric susceptibility have been observed \cite{Kovacs_dielectric_1997} after an electric filed perturbation but never after a mechanical perturbation. This new observation is indeed very useful because the measure of $\tnd$ as a function of frequency allows us to have information on the evolution of the DRT during the stretching, the waiting and the recovery intervals. The result of this measure strengthen previous observations on DRT \cite{Ediger2009} and brings new insight to the long lasting debate on mechanical rejuvenation in polymers (see for example \cite{McKenna_2003,Sahli_PRM_2020}). The observed time scaling and DRT evolution is also useful when one tries to fit this observation in the more general models of memory effects which have been developed to explain the experimental observation in glassy and disordered systems.   
%As a conclusion, we have shown that a polymer stretched at constant $\dgam$ presents complex memory effects after that  $\dgam$ has been set at zero  for a time $t_w$ at a given $\lambda_w$. We have measured the response of  the stress and of the dielectric constant. When  $\dgam$ is resumed  both observables present Kovacs-like non monotonic behavior in which the overshoot amplitude and position depend on $\Delta t/\log(t_w)$. However the recovery time to reach the reference curve,  is longer for  $\tnd$  than for $\sigma$. It is important to point out that the measure of $\tnd$ as a function of frequency during all the experiment allows us to have an information on the evolution of the DRT which plays an important role in  aging models. 
Thus our results, by showing important features of the dynamical processes of two interrelated quantities and of
the polymer relaxation dynamics, challenge the theoretical models of the memory effect in glassy materials and disordered systems.}

\acknowledgments 
J. Hem acknowledges the funding by Solvay. We thank O.Razebassia and A. Petrosyan for helpful assistance on the tensile machine and the Bartolo team for lending the cutting plotter. 

\bibliographystyle{apsrev4-1}
\bibliography{memory}

\begin{thebibliography}{56}
\bibitem [{nec()}]{necking_note}%
  \BibitemOpen
  \href@noop {} {}\bibinfo {note} {{\color{black} The sample deformation is
  homogeneous and there is no necking till the sample failure which occurs for
  $\lambda \ge 1.8$.}}\BibitemShut {Stop}%
\end{thebibliography}

\end{document}